\documentclass[prb,amsmath,amssymb,twocolumn]{revtex4}
\usepackage{amssymb,amsbsy,bbold,mathtools,relsize,bbold,mathrsfs}
\usepackage{amsthm}
\usepackage[bbgreekl]{mathbbol}

\pdfoutput=1
\allowdisplaybreaks

\begin{document}
\title{Magnetic Nernst effect}
\author{Sylvain D. \surname{Brechet}}
\email{sylvain.brechet@epfl.ch}
\author{Jean-Philippe \surname{Ansermet}}
\affiliation{Institute of Condensed Matter Physics, Station 3, Ecole Polytechnique F\'ed\'erale de Lausanne - EPFL, CH-1015 Lausanne,
Switzerland}

\begin{abstract}
	
The thermodynamics of irreversible processes in continuous media predicts the existence of a Magnetic Nernst effect that results from a magnetic analog to the Seebeck effect in a ferromagnet and magnetophoresis occurring in a paramagnetic electrode in contact with the ferromagnet. Thus, a voltage that has DC and AC components is expected across a Pt electrode as a response to the inhomogeneous magnetic induction field generated by magnetostatic waves of an adjacent YIG slab subject to a temperature gradient. The voltage frequency and dependence on the orientation of the applied magnetic induction field are quite distinct from that of spin pumping.

\end{abstract}

\maketitle

\section{Introduction}

In spincaloritronics, there has been recently quite some interest in the study of the propagation of spin waves across a ferromagnetic film in the presence of a temperature gradient~\cite{Cunha:2013,Hillebrands:2012}. For a configuration where the external magnetic induction field is parallel to the temperature gradient, the propagation of magnetization waves induces a magnetic induction field of magnitude proportional to the temperature gradient. Since the Seebeck effect refers to an electric field induced by a temperature gradient, this effect demonstrated in a YIG slab~\cite{Brechet:2013b} can be called the Magnetic Seebeck effect. 

The Magnetic Seebeck effect is a dynamical effect resulting from the precession of the magnetization. Thus, it should not be confused with the Spin Seebeck effect~\cite{Uchida:2008,Bosu:2011,Jaworski:2010,Uchida:2010,Uchida:2010b} where the magnetization is at equilibrium. A theoretical model of the Spin Seebeck effect was established by Adachi \textit{et al.}~\cite{Adachi:2012} in a quantum framework, while Schreier \textit{et al.}~\cite{Schreier:2013} attribute the effect to a difference in the temperatures of the lattice and the magnetization.

Here, we point out that the thermodynamics of irreversible processes in a magnetic continuous medium~\cite{Brechet:2013} predicts that the electrostatic potential depends on the gradient of the magnetic induction field applied to it. This is a consequence of the magnetophoretic force exerted on a magnetized charge carrier. Magnetophoresis is commonly used in electrochemistry~\cite{Leventis:2001} and biophysics~\cite{Furlani:2007}. The Magnetic Seebeck effect implies the existence of a magnetic induction field normal to the interface between the ferromagnet and the electrode. Thus, the magnetophoretically induced electric field, the thermally induced magnetic induction field and the temperature gradient are orthogonal to one another as in the Nernst effect. Hence, we call it the Magnetic Nernst effect~\cite{Callen:1960}. Here, the YIG ferromagnet is an insulator and the Pt electrode is a paramagnetic conductor. The Magnetic Nernst effect is thus a combination of the Magnetic Seebeck effect presented in reference~\cite{Brechet:2013b} and magnetophoresis~\cite{Lim:2011}. It is to be distinguished from the anomalous Nernst~\cite{Weiler:2012} effect and from the planar Nernst effect~\cite{Schmid:2013}.

As shown below, the composition of these two effects results in a voltage that has a DC component and an AC component oscillating at twice the frequency of the magnetization. The maximum amplitude occurs when the magnetic induction field applied to carry the ferromagnetic resonance is oriented with a $45^{\circ}$ angle with respect to the orientation of the electrode and of the temperature gradient.

\section{Magnetic Seebeck effect}

As shown below, the composition of these two effects results in a voltage that has a DC component and an AC component oscillating at twice the frequency of the magnetization. The maximum amplitude occurs when the magnetic induction field applied to carry the ferromagnetic resonance is oriented with a $45^{\circ}$ angle with respect to the orientation of the electrode and of the temperature gradient.

For the sake of clarity, we recall here in what sense a magnetic induction field is induced by an out-of-equilibrium magnetization in a temperature gradient. The formalism presented in reference~\cite{Brechet:2013} implies the existence of a magnetic counter-part to the well-known Seebeck effect, where a magnetic induction field $\boldsymbol{B}_{\,T}$ is induced by a temperature gradient $\boldsymbol{\nabla}_{y}\,T$ in a YIG slab in the presence of an oscillating magnetic excitation field and a constant external magnetic induction field $\boldsymbol{B}_{\,\text{ext}}\,$ in the slab plane (see axes $\boldsymbol{\hat{x}}$, $\boldsymbol{\hat{y}}$, $\boldsymbol{\hat{z}}$ on Fig.~\ref{YIG}). The existence of a magnetic induction field $\boldsymbol{B}_{\,T}$ can be understood as follows. In an insulator like YIG, there is no drift current. Thus, an induced magnetization force density $\boldsymbol{M}_{Y}\,\boldsymbol{\nabla}_{y}\,\boldsymbol{B}_{\,T}$ balances the thermal force density $-\,n_{Y}\,k_B\,\boldsymbol{\nabla}_{y}\,T$, i.e.
\begin{equation}\label{lin phen}
\boldsymbol{M}_{Y}\,\boldsymbol{\nabla}_{y}\,\boldsymbol{B}_{\,T} = \lambda_{Y}\,n_{Y}\,k_B\,\boldsymbol{\nabla}_{y}\,T\,,
\end{equation}
where the index $\vphantom{a}_Y$ refers to YIG, $\lambda_{Y}>0$ is a phenomenological dimensionless parameter, $\boldsymbol{M}_{Y}$ is the magnetization of YIG and $n_{Y}$ is the Bohr magneton number density of YIG. The magnetization $\boldsymbol{M}_{Y}$ is the sum of the saturation magnetization and the magnetic linear response, i.e. 
\begin{equation}\label{M_Y}
\boldsymbol{M}_{Y} = \boldsymbol{M}_{S_{Y}} + \boldsymbol{m}_{Y}\quad\text{where}\quad\boldsymbol{m}_{Y}\cdot\boldsymbol{B}_{\,\text{ext}}=0\,. 
\end{equation}
As detailed in reference~\cite{Brechet:2013}, the magnetic induction field $\boldsymbol{B}_{\,T}$ induced by the temperature gradient $\boldsymbol{\nabla}_{y}\,T$ can be written as,
\begin{equation}\label{B T}
\boldsymbol{B}_{\,T} = \boldsymbol{\varepsilon}_{\boldsymbol{M}_Y}\times\boldsymbol{\nabla}_{y}\,T\,,
\end{equation}
where the phenomenological vector $\boldsymbol{\varepsilon}_{\boldsymbol{M}_Y}$ is given by,
\begin{equation}\label{epsilon}
\boldsymbol{\varepsilon}_{\boldsymbol{M}_Y} = -\,\frac{\lambda_{Y}\,n_{Y}\,k_B}{M_{S_{Y}}^{2}}\,\left(\boldsymbol{\nabla}^{-1}_{y}\times\boldsymbol{m}_{Y}\right) \,.
\end{equation}
Hence, the thermally induced magnetic induction field $\boldsymbol{B}_{\,T}$ is oscillating at the frequency of $\boldsymbol{m}_{Y}\,$.

\section{Magnetophoresis}

The continuity of the orthogonal component of the thermally induced magnetic induction field $\boldsymbol{B}_{\,T}$ across the junction between the YIG and the Pt is ensured by Thomson's equation, i.e.
\begin{equation}\label{thomson}
\boldsymbol{\nabla}\cdot\boldsymbol{B}_{\,T} = 0\,.
\end{equation}
Therefore, the normal component of the magnetic induction field $\boldsymbol{B}_{\,T}$ is acting also on the Pt electrode. Magnetophoresis occurs in a Pt electrode that is sufficiently thick to be treated thermodynamically and sufficiently narrow for the temperature gradient to be neglected. The interaction between the magnetization of the conduction electrons of the paramagnetic Pt electrode and the thermally induced magnetic induction field in the ferromagnetic YIG slab results in a magnetization force that leads to the diffusion of the conduction electrons along the electrode, i.e. magnetophoresis. This generates in turn an electrostatic potential gradient $\boldsymbol{\nabla}_{x}\,V\,$ across the Pt electrode (see Fig.~\ref{YIG}) which can be thought of as a magnetophoretic electrochemical voltage. The existence of an the electrostatic potential gradient orthogonal to the temperature gradient depends on the orientation of the external magnetic induction field, as we shall show. 

Concretely, in the Pt electrode, the thermodynamic formalism~\cite{Brechet:2013} yields linear phenomenological relations between the electric current density and the magnetization and electrostatic forces densities respectively. By identifying the electric current in these relations, the electrostatic force density $-\,q_{P}\,\boldsymbol{\nabla}_{x}\,V$ resulting from the drift of the conduction electrons is found to be proportional to the magnetization force density $\boldsymbol{M}_{P}\,\boldsymbol{\nabla}_{x}\,\boldsymbol{B}_{\,T}$ generating the drift, i.e.
\begin{equation}\label{lin phen bis}
-\,q_{P}\,\boldsymbol{\nabla}_{x}\,V = \lambda_{P}\,\boldsymbol{M}_{P}\boldsymbol{\nabla}_{x}\,\boldsymbol{B}_{\,T}\,,
\end{equation}
where the index $\vphantom{a}_P$ refers to Pt, $q_{P}<0$ is the charge density of conduction electrons, $\boldsymbol{M}_{P}$ is the magnetization of the conduction electrons in the Pt electrode and $\lambda_{P}>0$ is a phenomenological dimensionless parameter. The magnetization $\boldsymbol{M}_{P}$ is the sum of the paramagnetic contribution due to the constant external field $\boldsymbol{B}_{\,\text{ext}}$ and the linear response $\boldsymbol{m}_{P}$ to the stray magnetic induction field generated by the propagating magnetization waves in the ferromagnet, i.e.
\begin{equation}\label{M_P}
\boldsymbol{M}_{P} = \frac{\chi_{P}}{\mu_{0}}\,\boldsymbol{B}_{\,\text{ext}} + \boldsymbol{m}_{P}\quad\text{where}\quad\boldsymbol{m}_{P}\cdot\boldsymbol{B}_{\,\text{ext}}=0\,, 
\end{equation}
$\mu_{0}$ is the magnetic permeability of vacuum, $\chi_{P}$ is the Pauli susceptibility of conduction electrons in Pt. As shown in reference~\cite{Griffiths:1999}, the magnetization force density can be expressed in terms of the magnetization current density, i.e.
\begin{equation}\label{Laplace force bis}
\boldsymbol{M}_{P}\,\boldsymbol{\nabla}_{x}\,\boldsymbol{B}_{\,T} = \left(\boldsymbol{\nabla}_{x}\times\boldsymbol{m}_{P}\right)\times\boldsymbol{B}_{\,T}\,.
\end{equation}
Thus, the relations~\eqref{lin phen bis} and~\eqref{Laplace force bis} imply that the electrostatic potential gradient generated by transport of the conduction electrons is given by,
\begin{equation}\label{V induced}
\boldsymbol{\nabla}_{x}\,V = -\,\frac{\lambda_{P}}{q_{P}}\,\Big(\left(\boldsymbol{\nabla}_{x}\times\boldsymbol{m}_{P}\right)\times\boldsymbol{B}_{\,T}\Big)\,.
\end{equation}
This effect is shown on Fig.~\ref{YIG} for a YIG slab with a Pt electrode.

\section{Magnetic Nernst effect}

\begin{figure}[htb]\hspace{3mm}
\begin{center}
\includegraphics[width=\columnwidth]{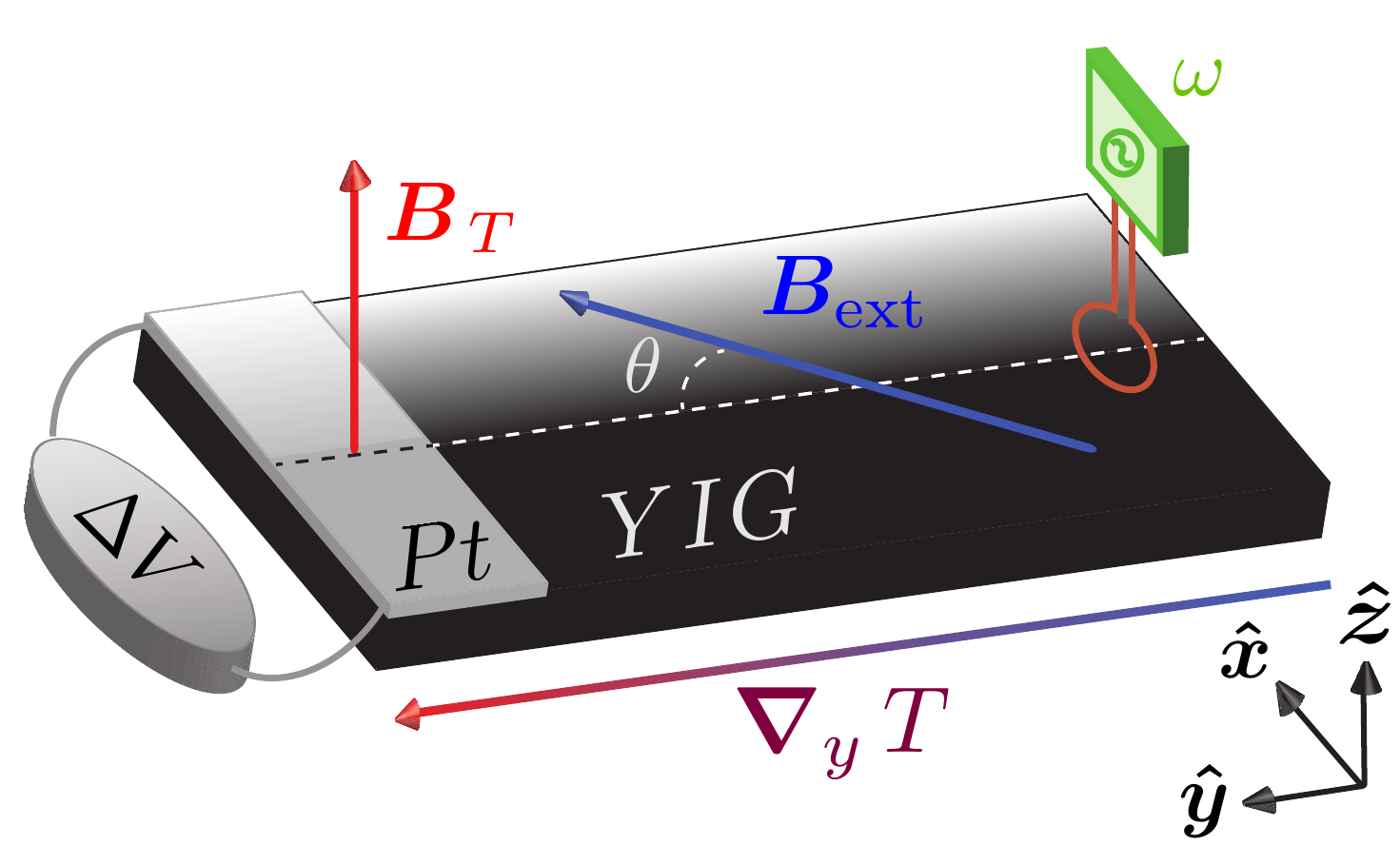}
\caption{YIG slab with a Pt electrode connected to a voltmeter and excited by a local probe.}
\label{YIG}
\end{center}
\end{figure}

The voltage difference derived from $\boldsymbol{\nabla}_{x}\,V$ in equation~\eqref{V induced} is a Magnetic Seebeck effect detected electrically through the magnetophoresis of the conduction electrons in the Pt electrode. We show now explicitly this effect in the form of a Nernst effect. In a sense the Magnetic Nernst effect is a thermally induced magnetophoresis. In the Magnetic Seebeck effect, the temperature gradient $\boldsymbol{\nabla}_{y}\,T$ imposed on the YIG slab induces a magnetic induction field $\boldsymbol{B}_{\,T}$ that is oscillating in an orthogonal plane. Through magnetophoresis, this field generates an electrostatic potential gradient $\boldsymbol{\nabla}_{x}\,V\,$ across the Pt electrode. Using the Magnetic Seebeck effect~\eqref{B T} and the definition~\eqref{epsilon}, the electrostatic potential gradient generated by magnetophoresis~\eqref{V induced} is recast explicitly as,
\begin{equation}\label{V induced bis}
\boldsymbol{\nabla}_{x}\,V = \gamma_{PY}\,\left(\boldsymbol{\nabla}_{x}\times\boldsymbol{m}_{P}\right)\times\Big(\left(\boldsymbol{\nabla}^{-1}_{y}\times\boldsymbol{m}_{Y}\right)\times\boldsymbol{\nabla}_{y}\,T\Big)\,,
\end{equation}
where
\begin{equation}\label{gamma}
\gamma_{PY} = \frac{\lambda_{P}\,\lambda_{Y}\,n_{Y}\,k_B}{q_{P}\,M_{S_{Y}}^2}\,.
\end{equation}
Using the Jacobi identity for the cross product, the linear relation~\eqref{V induced bis} yields the Magnetic Nernst effect, i.e.
\begin{equation}\label{V induced ter}
\boldsymbol{\nabla}_{x}\,V = \boldsymbol{N}_{\boldsymbol{m}}^{z}\times\boldsymbol{\nabla}_{y}\,T\,,
\end{equation}
where the phenomenological vector $\boldsymbol{N}_{\boldsymbol{m}}^{z}$ is given by,
\begin{equation}\label{Nernst N}
\boldsymbol{N}_{\boldsymbol{m}}^{z} = \gamma_{PY}\,\left(\boldsymbol{\nabla}_{x}\times\boldsymbol{m}_{Pz}\right)\times\left(\boldsymbol{\nabla}^{-1}_{y}\times\boldsymbol{m}_{Yz}\right)
\end{equation}
with $\boldsymbol{m}_{Pz}=\left(\boldsymbol{\hat{z}}\cdot\boldsymbol{m}_{P}\right)\boldsymbol{\hat{z}}$ and $\boldsymbol{m}_{Yz}=\left(\boldsymbol{\hat{z}}\cdot\boldsymbol{m}_{Y}\right)\boldsymbol{\hat{z}}\,$ in order to satisfy the vectorial symmetries and contribute to the effect. The structure of equation~\eqref{V induced ter} relating the gradients $\boldsymbol{\nabla}_{y}\,T$ and $\boldsymbol{\nabla}_{x}\,V$ is that of a Nernst effect. In place of a magnetic induction field, there is a phenomenological vector $\boldsymbol{N}_{\boldsymbol{m}}^{z}$, which depends on the out of equilibrium magnetization in the YIG slab. This effect is illustrated on Fig.~\ref{YIG} for a YIG slab with a Pt electrode, and a surface coil is presumed to excite the ferromagnetic resonance.

In order to determine the structure of the Magnetic Nernst vector $\boldsymbol{N}_{\boldsymbol{m}}^{z}$, we perform a Fourier series expansion of the linear response fields $\boldsymbol{m}_{Pz}$ and $\boldsymbol{m}_{Yz}$. In a stationary regime, the Fourier transform of the response fields are expressed in terms of real parameters as,
\begin{align}
\label{m P}
&\boldsymbol{m}_{Pz} = \sum_{\boldsymbol{k}_{P}}\,m_{\boldsymbol{k}_{P}z}\,\sin\left(\boldsymbol{k}_{P}\cdot\boldsymbol{r}-\,\omega_{\boldsymbol{k}_{P}} t + \phi_{\boldsymbol{k}_{P}}\right)\,\boldsymbol{\hat{z}}\,,\\
\label{m Y}
&\boldsymbol{m}_{Yz} = \sum_{\boldsymbol{k}_{Y}}\,m_{\boldsymbol{k}_{Y}z}\,\sin\left(\boldsymbol{k}_{Y}\cdot\boldsymbol{r}-\,\omega_{\boldsymbol{k}_{Y}} t + \phi_{\boldsymbol{k}_{Y}}\right)\,\boldsymbol{\hat{z}}\,,
\end{align}
where $\phi_{\boldsymbol{k}}$ and $\varphi_{\boldsymbol{k}}$ are the dephasing angles and $\omega_{\boldsymbol{k}}$ is the angular frequency of the eigenmodes $\boldsymbol{k}$. Finally, the Fourier decompositions~\eqref{m P} and~\eqref{m Y} imply that the relation~\eqref{Nernst N} is recast in terms of the eigenmodes as,
\begin{align}\label{Nernst N bis}
&\boldsymbol{N}_{\boldsymbol{m}}^{z} = \gamma_{PY}\!\!\sum_{\boldsymbol{k}_{P},\boldsymbol{k}_{Y}}\!k_{P_x}\,k_{Y_y}^{-1}\,m_{\boldsymbol{k}_{P}z}\,m_{\boldsymbol{k}_{Y}z}\\
&\!\cdot\cos\left(\boldsymbol{k}_{P}\!\cdot\!\boldsymbol{r}-\omega_{\boldsymbol{k}_{P}} t + \phi_{\boldsymbol{k}_{P}}\!\right)\cos\left(\boldsymbol{k}_{Y}\!\cdot\!\boldsymbol{r}-\omega_{\boldsymbol{k}_{Y}} t + \phi_{\boldsymbol{k}_{Y}}\!\right)\boldsymbol{\hat{z}}\,,\nonumber
\end{align}
where $k_{P_{x}} = \boldsymbol{\hat{x}}\cdot\boldsymbol{k}_{P}$ and $k_{Y_{y}}^{-1} = \boldsymbol{\hat{y}}\cdot\boldsymbol{k}_{Y}^{-1}\,$. The magnetic waves vectors in YIG and Pt are collinear to the external magnetic induction field, i.e. $\boldsymbol{k}_{P}\times\boldsymbol{B}_{\,\text{ext}}=\boldsymbol{0}$ and $\boldsymbol{k}_{Y}^{-1}\times\boldsymbol{B}_{\,\text{ext}}=\boldsymbol{0}$, which implies that $k_{P_{x}} = k_{P}\,\sin\theta$ and $k_{Y_{y}}^{-1} = k_{Y}^{-1}\,\cos\theta$ where $\theta$ is the orientation angle between the temperature gradient and the external magnetic field in the plane of the YIG slab as shown on Fig.~\ref{YIG}. Moreover, the specific mode $\boldsymbol{k}=\boldsymbol{k}_{Y}=\boldsymbol{k}_{P}$ corresponding to the excitation frequency $\omega\equiv\omega_{\boldsymbol{k}}$ of the magnetization waves in YIG and Pt is determined by the quadratic dispersion relation of the magnetization waves , i.e. $\omega_{\boldsymbol{k}} = A \boldsymbol{k}^2$ where $A$ is the stiffness. Thus, choosing the initial time to cancel the dephasing of the magnetization in YIG and using the trigonometric identity,
\begin{equation}\label{trigo}
\begin{split}
&\cos\left(\boldsymbol{k}\cdot\boldsymbol{r}-\,\omega t + \phi\right)\cos\left(\boldsymbol{k}\cdot\boldsymbol{r}-\,\omega t\right) =\\
&\frac{1}{2}\,\Big(\cos\phi +\cos\left(2\boldsymbol{k}\cdot\boldsymbol{r}-\,2\omega t + \phi\right)\Big)\,,
\end{split}
\end{equation}
the Magnetic Nernst vector~\eqref{Nernst N bis} is recast as,
\begin{equation}\label{Nernst N ter}
\begin{split}
&\boldsymbol{N}_{\boldsymbol{m}}^{z} = \frac{\gamma_{PY}}{4}\ m_{\boldsymbol{k}_{P}z}\,m_{\boldsymbol{k}_{Y}z}\,
\sin\left(2\theta\right)\\
&\cdot\Big(\cos\phi + \cos\left(2\boldsymbol{k}\cdot\boldsymbol{r} -\,2\omega t + \phi\right)\Big)\,\boldsymbol{\hat{z}}\,,
\end{split}
\end{equation}
where $\phi\equiv\phi_{\boldsymbol{k}_P}$ and $\phi_{\boldsymbol{k}_Y}=0\,$. In the homogeneous electrode, the electrostatic potential varies linearly along the $\boldsymbol{\hat{x}}$-axis, i.e. $\boldsymbol{\hat{x}}\cdot\boldsymbol{\nabla}\,V = \Delta V/\ell_{x}$ where $\ell_{x}$ is the length of the electrode. Thus, the voltage across the electrode is given by,
\begin{equation}\label{V induced quad}
\Delta\,V = \ell_{x}\,\boldsymbol{\hat{x}}\cdot\left(\boldsymbol{N}_{\boldsymbol{m}}^{z}\times\boldsymbol{\nabla}_{y}\,T\right)\,.
\end{equation}
The expressions~\eqref{Nernst N ter} and~\eqref{V induced quad} imply that the voltage $\Delta\,V$ along the electrode consists of DC and AC contributions. The DC contribution is proportional to $\cos\phi$, which implies that it is maximal in the absence dephasing between the magnetization in Pt and YIG. The AC contribution is oscillating with an angular frequency $2\omega$ that corresponds to the double of the excitation angular frequency $\omega$. The Magnetic Nernst effect vanishes if the external magnetic field $\boldsymbol{B}_{ext}$ is collinear ($\theta=0$) or orthogonal ($\theta=\pi/2$) to the temperature gradient $\boldsymbol{\nabla}_{y}\,T$ and it is maximal if there is a $\pi/4$ angle between these vectors in the YIG slab plane.

It is important to mention that the Magnetic Nernst effect is not equivalent to thermal spin pumping~\cite{Uchida:2012}. The angular dependence of these two effects are different. Thermal spin pumping is maximal for $\theta=0$ and minimal for $\theta=\pi/2$ whereas the Magnetic Nernst effect is maximal for $\theta=\pi/4$ and minimal for $\theta=0$ and $\theta=\pi/2\,$.

\section{Conclusion}

In summary, a Nernst effect is predicted, which results from the interplay between the magnetization dynamics of a ferromagnet driven in a temperature gradient, and the linear response of the paramagnetic electrode to the inhomogeneous field produced by the magnetization.

First, in a magnetic insulating YIG slab, there is the Magnetic Seebeck effect, i.e. a magnetic induction field induced by a temperature gradient. Second, the electrical detection of this effect in a paramagnetic Pt electrode contacted to the slab relies on the voltage induced by the inhomogeneous field acting on the conductive electrode, causing a drift of charges that carry a magnetic dipole. 

The voltage has a DC component and an AC component that oscillates at twice the frequency of the field driving the magnetization. The effect is zero when the applied field is parallel or perpendicular to the temperature gradient, and maximum at a $45^{\circ}$ angle in between. Hence, this effect is quite distinct from the angular dependence of spin pumping or the Spin Seebeck effect.

\begin{acknowledgments}

We would like to thank Fran\c{c}ois Reuse and Klaus Maschke for useful comments and acknowledge the following funding agencies : Polish-Swiss Research Program NANOSPIN PSRP-$045/2010$; Deutsche Forschungsgemeinschaft SS$1538$ SPINCAT, no. AN$762/1$.

\end{acknowledgments}

\bibliography{references}

\end{document}